\documentclass[11pt,letterpaper]{article}
\usepackage[utf8]{inputenc}
\usepackage{amsmath, amsfonts, amssymb}
\usepackage{verbatim}
\usepackage{graphicx}
\usepackage{mathrsfs}
\usepackage{cite}
\usepackage{physics}
\usepackage{hyperref}
\usepackage[margin=2.5cm]{geometry}
\usepackage{multicol}
\usepackage{setspace}
\usepackage{secdot}
\usepackage{caption}
\usepackage{subcaption}
\usepackage{indentfirst}

\title{\bfseries Critical collapse for the Starobinsky $R^2$ model}
\author{
    Yolbeiker Rodr\'iguez Baez$^{\dagger,}\footnote{E-mail: yolbeiker.rodriguez@pucv.cl}$ \\
    \\\textit{$^\dagger$Pontificia Universidad Cat\'olica de Valpara\'\i so, Instituto de F\'\i sica,} 
    \\\textit{Av. Brasil 2950, Valpara\'{\i}so, Chile.}
    }
\date{December 2022}

\begin{document}
\maketitle
\begin{abstract}
We study gravitational collapse for the Starobinsky $R^2$ model, a particular example of an $f(r)$ theory, in a spherically symmetric spacetime. 
We add a massless scalar field as matter content to the spacetime. We work in the Einstein frame, where an additional scalar field arises due to the conformal transformation.
As in general relativity, depending on the initial data, we found that the gravity scalar field and the physical scalar field can collapse, forming a black hole, in which the final solution is the Schwarzschild metric. We found the threshold of black hole formation through a fine-tuning method and studied critical collapse near this regime. 
\end{abstract}
\spacing{1.3}

\section{Introduction}

Einstein’s theory of general relativity (GR) has been our best classical theory to describe gravity. All direct experimental tests are consistent with the theory \cite{Will:2014kxa}. However, convincing evidence tells us that GR is not the final theory to describe gravity. Many arguments indicate that GR is not renormalizable in the standard quantum field theory sense \cite{Stelle:1976gc, Biswas:2011ar, Biswas:2013cha}. 
It is geodesically incomplete for most of its solutions, like black holes that contain spacetime singularity, and does not explain many inconsistencies in the early Universe \cite{Hawking:1973uf}.

Many modifications of the theory have been considered in the literature. The most usual modifications consist of adding new degrees of freedom, by adding new fundamental fields in the action, or considering a spacetime with higher dimensions \cite{Berti:2015itd}. 
Another alternative is to modify the scalar curvature term in the Einstein-Hilber action. 
In this modification, known as $f(R)$ gravity theories \cite{Carroll:2003wy, Hu:2007nk, Starobinsky:2007hu}, the Ricci scalar, $R$, in the Einstein-Hilbert action is replaced by an arbitrary function of $R$.  
The action reads
\begin{equation}
    \mathcal{S}=\frac{1}{16\pi G}\int \dd^{4}x \sqrt{-g}f(R) + S_{m}, 
\end{equation}
where $G$ is the Newtonian constant, and $S_m$ is the action describing the matter content in spacetime.
As shown, $f(R)$ gravity is a natural extension of GR; see Refs. \cite{Sotiriou:2008rp, DeFelice:2010aj} for reviews of this theory.

The above action is written in the so-called Jordan frame. As it is well known, making a conformal transformation to pass to the Einstein frame, $f(R)$ theory is equivalent to a scalar-tensor theory. But, it is of a unique type since the potential describing the interaction with the scalar degree of freedom is related to the Ricci scalar $R$. 
$f(R)$ theory has been studied in great detail from the cosmological point of view in order to explain the accelerated expansion of the universe \cite{Nojiri:2003ft, Nojiri:2003ni, Carroll:2003wy, Dobado:1994qp, delaCruz-Dombriz:2006kob, Nojiri:2006ri} and the dark matter origin \cite{Cembranos:2008gj}.

In this paper, we are interested in studying the formation of black holes in $f(R)$ theory. 
It is well known that astrophysical black holes are formed by the gravitational collapse of matter. 
Black hole (BH) physics and spherical collapse have been key routes to understanding gravity. A remarkable property of a black hole is that only a few parameters entirely describe its gravitational field. This is the so-called no-hair conjecture, or the uniqueness theorem \cite{Chrusciel:2012jk}.

Modifying Einstein-Hilbert's action could give rise to new BH solutions and violation of the uniqueness theorem (see Ref. \cite{delaCruz-Dombriz:2009pzc}, and references therein, for static and spherically
symmetric black hole solutions in $f(R)$ gravity).
It was shown in Ref. \cite{WHITT1984176} that the fourth-order gravity theory with lagrangian $R - 2\Lambda + \alpha R^2$ is conformally equivalent to Einstein gravity with a massive scalar field. Being the Schwarzschild solution, the only static spherically symmetric solution. 

Understanding gravitational collapse phenomena have been a critical point in understanding the nature of the gravitational field. Based on GR, the dynamics of an isolated system that interacts only through the gravitational field typically end up either with the formation of a single black hole or a complete dispersion of the mass-energy to infinity.
The numerical study of spherical gravitational collapse has a long history. 
However, one of the pioneering works in this area was performed by M. Choptuik \cite{Choptuik:1992jv}, who solved the dynamical evolution of a scalar field minimally coupled to gravity from the numerical point of view. His investigations were inspired by previous analytical studies made by D. Christodoulou \cite{Christodoulou86-a, Christodoulou86-b, Christodoulou87}. 

Choptuik \cite{Choptuik:1992jv} has shown that if $p$ is a parameter describing some aspect of the initial distribution of scalar field energy, there exists a critical value $p^\star$ which denotes the threshold of black hole formation. 
For $p<p^\star$, the scalar field disperses to infinity, while for $p>p^\star$ a black hole is formed. In the supercritical regime, meaning for $p>p^\star$ but very close to the threshold, a universal behavior appears (i.e., independent of the initial data) relating the mass $M$ of black holes to a universal scaling behavior
\begin{align}
M\propto (p-p^\star)^{\gamma}~,~~~\gamma\simeq 0.37 \, .
\end{align}
This solution has been repeatedly verified by using a fully 3D code \cite{Healy:2013xia}.
Adding a mass term to the theory \cite{Brady:1997fj}, which introduces a length scale, also produces a universal behavior but with a different scaling parameter $\gamma$.

The pioneering work of Choptuik opened a new line of investigation to study the gravitational collapse of other fields. 
For example, the previous work has been extended to massive scalar fields \cite{Brady:1997fj, Jimenez-Vazquez:2022fix},
the gravitational collapse of radiation \cite{Evans:1994pj} and perfect fluid \cite{Neilsen00}, a complex scalar field with angular momentum \cite{PhysRevLett.93.131101}, K-essence models \cite{Gannouji:2020kas}, the Einstein-Massless-Dirac system \cite{Ventrella:2003fu}, and massive vector field \cite{Garfinkle:2003jf}, to mention some of the many investigations carried out to date. 

The studies on the gravitational collapse in $f(R)$ gravity have been studied in some particular cases \cite{Casado-Turrion:2022xkl}. For example, spherical scalar collapse in the Hu-Sawicki model, Starobinsky $R^2$ model, and $R\ln R$ model was studied in Refs. \cite{Guo:2013dha, Guo:2015eho}. In these papers, they employed double-null coordinates in the Einstein frame, focusing the analysis on the dynamics in the vicinity of the singularity of the formed black hole.
Fluid gravitational collapse in $f(R)$ gravity was examined in \cite{Sharif:2010um, Sharif:2013boa, Sharif:2014tqa, Kausar:2014uda, Chakrabarti:2016teb, Astashenok:2018bol}. 
The gravitational collapse of a charged black hole in $f(R)$ gravity was studied in \cite{Hwang:2011kg} by using double-null formalism and the mass inflation of the Cauchy horizon was examined. 
A $f(R)$ model with a uniformly collapsing cloud of self-gravitating dust particles is studied in \cite{Cembranos:2012fd}. This analysis shares analogies with the formation of large-scale structures in the early Universe and with the formation of stars in a molecular cloud experiencing gravitational collapse. 
 Dark matter halo formation was studied in \cite{Kopp:2013lea}. 
The gravitational collapse of massive stars in $f(R)$ gravity was analyzed in \cite{Goswami:2014lxa}. 

In this paper, we study the spherically symmetric gravitational collapse of a massless scalar matter field in asymptotic flat spacetime in the Starobinsky $R^2$ gravity, one specific model in the $f(R)$ gravity.
A similar analysis was performed in \cite{Zhang:2016kzg}, but we extend the previous analysis by exploring the critical phenomena in the vicinity of the formation of the apparent horizon.

The organization of the paper is as follows: In section \ref{sec:Framework}, we will derive the equations of motion for a general $f(R)$ theory. We start from the Jordan frame and make the conformal transformation to write this theory as a Scalar-Tensor theory. In section \ref{sec:metric_equations}, we assume a spherically symmetric spacetime and derive the equations for gravitational collapse. Then in section \ref{sec:numerical_method}, we will explain the numerical method. We will report
numerical results in section \ref{sec:results}. In the final section, we will present conclusions and discussions.

\section{Framework of $f(R)$ gravity \label{sec:Framework}}

This section briefly summarized the equations derived from the $f(R)$ theory. 
We start with the action written in the Jordan frame \cite{Sotiriou:2008rp, DeFelice:2010aj}
\begin{equation} \label{fR_action} 
    \mathcal{S}_J = \frac{1}{16\pi G}\int \dd^{4}x \sqrt{-g}f(R) + S_{m}, 
\end{equation}
where $G$ is the Newtonian constant, $g$ is the determinant of the metric $g_{\mu\nu}$, and $S_{m}$ describes the matter content in the spacetime.
Variation of the action with respect to the metric tensor gives us
\begin{equation}\label{fR_Equations} 
    f'R_{\mu\nu}-\frac{1}{2}f g_{\mu\nu} -\left(\nabla_{\mu}\nabla_{\nu}-g_{\mu\nu} \Box\right) f'
= 8\pi G \, T^{(M)}_{\mu\nu},
\end{equation}
where $f' \equiv \partial_R f(R)$ denotes the derivative of the function $f$ with respect to its argument $R$, $\Box$ is the usual
notation for the covariant D'Alembert operator $\Box\equiv\nabla_{\alpha}\nabla^{\alpha}$, $\nabla$ is the covariant derivative, and $T^{(M)}_{\mu\nu}$ is the energy-momentum tensor for the matter field. 
Defining
\begin{equation}
    \chi \equiv \frac{\partial f(R)}{\partial R} \, , \qquad 
    U'(\chi)\equiv\frac{dU}{d\chi}=\frac{1}{3}(2f-\chi R),
\end{equation}
one can write the trace of the Eq.(\ref{fR_Equations}) as 
\begin{equation}\label{trace_eq1} 
    \Box \chi=U'(\chi)+\frac{8\pi G}{3}T.
\end{equation}
Thus, $\chi = f'(R)$ is a new scalar degree of freedom that is not present in GR. 
In fact, for Einstein's gravity, $f(R) = R \Longrightarrow \chi\equiv1$ and $\Box \chi\equiv0$, the above equation reduces to the trace of the usual Einstein equations.

In the Jordan frame, the gravitational Lagrangian is an arbitrary function of $R$, and Einstein's equations are usually fourth-order in the metric.
In order to make the formalism less involved, we transform $f(R)$ gravity from the current frame into the Einstein frame.
In the latter, the second-order derivatives of $f'(R)$ are absent in the equations of motion for the metric components. 
We will see that the formalism can be treated as Einstein's gravity coupled to a scalar field. 
Therefore, we can use some results that have been developed in the numerical relativity community.

Taking the conformal transformation $\tilde{g}_{\mu\nu} = \Omega^2 g_{\mu\nu} $, where the conformal factor is given by
\begin{equation} \label{eq:conformal_trans}
   \Omega^2 = f'(R) = \exp\qty(\sqrt{\frac23} \kappa \phi) = \chi ,
\end{equation}
one obtains the corresponding action of $f(R)$ gravity in the Einstein frame
\begin{eqnarray}\label{eq:fR_action_EF}
\mathcal{S}_E = \int \dd^4 x\sqrt{-\tilde g}\left[\frac{1}{2\kappa^2} \tilde R -\frac{1}{2} \tilde g^{\mu\nu}\tilde\nabla_\mu\phi\tilde\nabla_\nu\phi-V(\phi)\right] + \int \dd^4 x\mathcal{L}_M(\tilde g_{\mu\nu} / {\chi(\phi)},\psi),
\end{eqnarray}
where $\kappa^2={8\pi G}$, and a tilde denotes quantities in the Einstein frame; thus, $\tilde\nabla$ and $\tilde R$ are the covariant derivative and the Ricci scalar associated to the metric $\tilde g_{\mu\nu}$. 
In this action, $\phi$ is a new scalar field that takes into account the modifications of Einstein gravity (in this paper, usually referred to as the gravity scalar field), the potential is given by
\begin{equation}\label{eq:fR_potential}
    V(\phi) \equiv \frac{f' R - f}{2\kappa^2 f'^2} \,,
\end{equation}
and $\psi$ is the matter field that, in the Einstein frame, is non-minimally coupled to gravity.

From the action (\ref{eq:fR_action_EF}), we can proceed as usual. Variation with respect to the metric field, $\tilde g_{\mu\nu}$, give us the Einstein field equations 
\begin{equation} \label{eq:Einstein_Eq}
    \tilde G_{\mu\nu} \equiv \tilde R_{\mu\nu} - \frac12 \tilde R \tilde g_{\mu\nu} 
    = \kappa^2 \qty[\tilde T^{(\phi)}_{\mu\nu} + \tilde T^{(M)}_{\mu\nu}] \, ,
\end{equation}
where we have defined the energy-momentum tensors as
\begin{align}
\tilde T_{\mu\nu}^{(\phi)} &= \partial_\mu\phi\partial_\nu\phi-\tilde g_{\mu\nu}\left[\frac{1}{2}\tilde g^{\alpha\beta}\partial_\alpha\phi\partial_\beta\phi+V(\phi)\right], 
\qquad
\tilde T_{\mu\nu}^{(M)} = \frac{T_{\mu\nu}^{(M)}}{\chi}.
\end{align}

We are interested in the gravitational collapse of the most simple type of matter, a massless scalar field $\psi$. 
The Lagrangian density is given by
\begin{equation} \label{eq:Lag_matter}
    \mathcal{L}_M = -\frac12 \sqrt{-g} \; g^{\mu\nu}\nabla_\mu \psi\nabla_\nu \psi ;
\end{equation}
thus, the energy-momentum tensor for the matter field, in the Einstein frame, is
\begin{align}
  \tilde T_{\mu\nu}^{(M)} & = 
  \frac{1}{\chi}\left(\partial_\mu \psi \partial_\nu \psi-\frac{1}{2}g_{\mu\nu}g^{\alpha\beta}\partial_\alpha\psi\partial_\beta \psi\right)
  = \frac{1}{\chi} \left(\partial_\mu \psi \partial_\nu \psi - \frac{1}{2}\tilde g_{\mu\nu} \tilde g^{\alpha\beta}\partial_\alpha\psi\partial_\beta \psi\right) \\
  & = \exp\qty(- \sqrt{\frac23} \kappa \phi) \left(\partial_\mu \psi \partial_\nu \psi - \frac{1}{2}\tilde g_{\mu\nu} \tilde g^{\alpha\beta}\partial_\alpha\psi\partial_\beta \psi\right) \,.
\end{align}

Using Eq.(\ref{eq:Lag_matter}), variation of the action (\ref{eq:fR_action_EF}) with respect to the scalar fields, gives us
\begin{align}
    \tilde\Box \phi & - \frac{\partial V(\phi)}{\partial\phi} - \frac{\kappa}{\sqrt{6}}  \tilde T^{(M)} = 0,  \label{fR:Box_phi} \\
    \tilde\Box \psi & - \sqrt{\frac{2}{3}} ~\kappa \tilde g^{\mu\nu} \partial_\mu \phi\partial_\nu \psi = 0,
\label{fR:Box_psi}
\end{align}
where $\tilde T^{(M)}$ is the trace of the energy-momentum tensor of the matter field.
For the massless scalar field, Lagrangian (\ref{eq:Lag_matter}), we obtain
$\tilde T^{(M)} = - \exp\qty(- \sqrt{\frac23} \kappa \phi) \tilde g^{\alpha\beta}\partial_\alpha\psi\partial_\beta \psi$.

Henceforth, we work only with equations of motion in the Einstein frame; thus, we drop the tilde over the quantities. 

\section{Metric ansatz and equations \label{sec:metric_equations}}

As mentioned before, we are interested in studying the formation of event horizon in $f(R)$ theories. 
We use the metric in polar-areal coordinates \cite{Choptuik:1992jv}, which takes the form
\begin{equation}\label{eq:choptuik_metric}
    \dd s^2 = -\alpha^2(t,r) \dd t^2 + a^2(t,r) \dd r^2 + r^2 (\dd \theta^2 + \sin^2\theta \dd\varphi^2) \, ,
\end{equation}
where, $\alpha$ and $a$ are the metric functions that depend on $time$ and the $radial$ coordinates in order to study the evolution.
Using this metric ansatz, the equations of motion for the scalar fields (\ref{fR:Box_phi}, \ref{fR:Box_psi}) are second-order PDEs. 
In order to reduce the field equations to a system of first-order PDEs, we define the following auxiliary fields \cite{Choptuik:1992jv, Zhang:2016kzg}
\begin{equation}
    \begin{aligned}
    Q(t,r) & = \partial_r \phi(t,r) \,, \qquad \qquad 
    P(t,r) = \frac{a}{\alpha} \partial_t  \phi(t,r) \, ,
    \\
    \Phi(t,r) & = \partial_r \psi(t,r) \,, \qquad \qquad 
    \Pi(t,r) = \frac{a}{\alpha} \partial_t  \psi(t,r) \, .
\end{aligned}
\label{eq:AuxFields_def}
\end{equation}

Thus, Eqs.(\ref{fR:Box_phi}) and (\ref{fR:Box_psi}) are written as\footnote{Eqs.(\ref{dt_QQ}) and (\ref{dt_Fi}) come from the definitions of the auxiliary fields, and Eqs.(\ref{dt_PP}) and (\ref{dt_Pi}) from the Klein-Gordon equations of the fields.}
\begin{align}
    \partial_t P & = \frac{1}{r^2}\frac{\partial }{\partial r}\qty(r^2 \frac{\alpha}{a}Q) - a \alpha \frac{\partial V(\phi)}{\partial \phi } - e^{-\sqrt{\frac{2}{3}} \phi} \frac{\kappa \alpha}{\sqrt{6} a} \left(\Pi ^2-\Phi ^2\right) \, , \label{dt_PP}
    \\
    \partial_t Q & = \frac{\partial }{\partial r}\qty(\frac{\alpha}{a} P) \, , \label{dt_QQ}
    \\
    \partial_t \Pi & = \frac{1}{r^2}\frac{\partial }{\partial r}\qty(r^2 \frac{\alpha}{a}\Phi) + \sqrt{\frac23} \frac{\kappa\alpha}{a} (P\Pi - Q\Phi) \, , \label{dt_Pi}
    \\
    \partial_t \Phi & = \frac{\partial }{\partial r}\qty(\frac{\alpha}{a} \Pi) \, . \label{dt_Fi}
\end{align}
To fix the evolution of the system, we need the equations for the metric variables, these are obtained from Eq.(\ref{eq:Einstein_Eq}). 
The $tt-$ and $rr-$components of Einstein equation gives
\begin{align}
    \frac{a'}{a} = \frac{1-a^2}{2 r}+\frac{\kappa^2}{4} r \left(e^{-\sqrt{\frac{2}{3}} \phi } \left(\Pi ^2+\Phi ^2\right)+P^2+Q^2\right)+\frac{\kappa^2}{2} r a^2 V(\phi) \, ,
    \\
\frac{\alpha'}{\alpha } = \frac{a^2-1}{2 r}+\frac{\kappa^2}{4}  r \left(e^{-\sqrt{\frac{2}{3}} \phi } \left(\Pi ^2+\Phi ^2\right)+P^2+Q^2\right)-\frac{\kappa^2}{2}  r a^2 V(\phi) \, ,
\end{align}
here, a prime denotes a derivative with respect to the radial coordinate. 
As we can see, these equations do not describe evolution in time; they are constraint equations that must be satisfied at each time.
Thus, the evolution of the metric functions at each hypersurface (at each time) is given by the values of the auxiliary fields in that hypersurface. 

As in GR, it can be checked that the system of PDEs (\ref{eq:Einstein_Eq}) is over-determined and can provide another equation for one of the metric functions. In fact, the $\theta\theta-$component of Eq. (\ref{eq:Einstein_Eq}) gives an evolution equation for metric variable $a$. We do not solve this equation, but we use it to monitor its convergence to zero as a check for the correctness of our solution. 

\section{Numerical setup \label{sec:numerical_method}}

In this section, we present the numerical formalisms that we used to solve numerically the system of PDEs, including boundary conditions, initial conditions, and how we detect a formation of a black hole.

\subsection{Initial Conditions}

The foliation of the spacetime allowed us to rewrite Einstein equations (\ref{eq:Einstein_Eq}) in a set of partial differential equations. To completely specify the Cauchy problem, we need to provide initial data to obtain exact solutions. We follow the same scheme that Choptuik used in \cite{Choptuik:1992jv}, in which a smooth function parameterizes the initial distribution of the scalar field. 

At initial time ($t=t_0=0$), we fix the form of each scalar field to be a Gaussian distribution
\begin{equation}\label{eq:init_data}
    \phi(0, r) = p_o^{(\phi)} \exp\qty[-\qty(\frac{r-r_o}{d})^2 ] \, , \qquad
    \psi(0, r) = p_o^{(\psi)} \exp\qty[-\qty(\frac{r-r_o}{d})^2 ] 
\end{equation}
where $\qty(p_o^{(\phi)}, \, p_o^{(\psi)}, \, r_o, \, d)$ are constant parameters. 
We keep only $\qty(p_o^{(\phi)}, \, p_o^{(\psi)})$ as free parameters, the others are fixed to $(r_0=20, \, d=3)$.
Once the initial data for both scalar fields are established, we are free to choose the initial data for the auxiliary fields (\ref{eq:AuxFields_def}). 
For example, we set
\begin{equation}
    \begin{aligned}
    \Phi(0, r) = \Pi(0, r) = \partial_r \psi(0, r)  \, , \\ 
    Q(0, r) = P(0, r) = \partial_r \phi(0, r)  \, ,
\end{aligned}
\end{equation}
which means that each scalar field is initially ingoing, $\phi$ and $\psi$ approach to the origin. 

\subsection{Boundary conditions}

In this paper, the range for the spatial coordinate is $r \in (0, \, r_{max})$; where $r_{max}$ simulates the infinity and numerically has to be chosen in a way that our solutions do not depend on this parameter. This can be done by imposing boundary conditions at the edges of this range.

The first boundary is located at $r = 0$, which is the origin of the spherically symmetric coordinate system.
We impose regularity conditions to ensure that fields at this point are not ill-defined. Therefore, analyzing the system of equations near $r\to 0$ and imposing spacetime to be locally flat
there, we enforce $\phi$ and $\psi$ to satisfy the following conditions:
\begin{equation}
    \partial_r \phi (0, r) = 0 \, , \qquad \partial_r \psi (0, r) = 0 \, .
\end{equation}
Also, the regularity of the system imposes $a(t, r = 0) = 1$.

Our second boundary is located at $r = r_{max}$. 
In this boundary, we must impose boundary conditions to have a zero amplitude for signals entering the integration domain. Therefore, we impose radiation boundary conditions \cite{ISRAELI1981115} at $r = r_{max}$. 
It can be mentioned that, in our numerical evolution, this boundary is relevant for those simulations in which there is no BH formation and the matter field is dispersed to infinity. We fix the value of $r_{max}$ in such a way that when a BH is formed, the reflected matter does not have enough time to reach the outer boundary.

\subsection{Black hole detection}

In the weak-field regime (when the amplitudes of the scalar fields are close to zero), it is expected that the matter content travels toward the origin and then, it is dispersed to infinity. 
As we increase the initial amplitude of the matter field $p_o^{(\phi)}$ or the initial amplitude of the gravity scalar field $p_o^{(\psi)}$, a black can be formed since we increase the amount of matter in spacetime and/or the gravity potential becomes stronger. 

Varying $\qty{p_o^{(\phi)}, \, p_o^{(\psi)}}$ parameters, we can reach the limit where a black hole forms in the evolution. However, as we stated before, our metric ---written in the polar/areal coordinates (\ref{eq:choptuik_metric})--- cannot penetrate apparent horizons. 

Nevertheless, we can detect signals that indicate the formation of a black hole: 
the lapse function goes to zero, signaling that the coordinate system starts to become singular near what would be the radius of the black hole.
In spherical symmetry, we can monitor the {\it Misner-Sharp mass} \cite{PhysRev.136.B571}
\begin{equation}
m_{MS}(t,r) = \frac{r}{2}\qty( 1 - (\nabla r)^2) = \frac{r}{2}\qty(1 - \frac{1}{a^2(t,r)}) \, .
\end{equation}

In a collapse simulation where a black hole is formed, we will see that the function $2m_{MS}/r$ rapidly tends to 1 at some specific radius,
$r = R_{BH}$, where $R_{BH}$ is precisely the radius of the black hole.
Numerically, we say that we have detected a black hole if $2m_{MS}/r \geq 0.995$.

\section{Model and results \label{sec:results}}

For numerical purposes, we need to specify the exact form of $f(R)$ gravity. 
Thus, we take the ansatz
\begin{equation} \label{eq::fR_model}
    f(R) = R + a R^2 
\end{equation}
where $a$ is a constant parameter; specifically, we chose $a=1$. 
This model is often called the Starobinsky $R^2$ model, which describes the inflation in the early universe. For details of the model, see \cite{Starobinsky:1980te} and the review \cite{DeFelice:2010aj, Vilenkin:1985md}.

\begin{figure}
    \centering
    \includegraphics[scale=0.8]{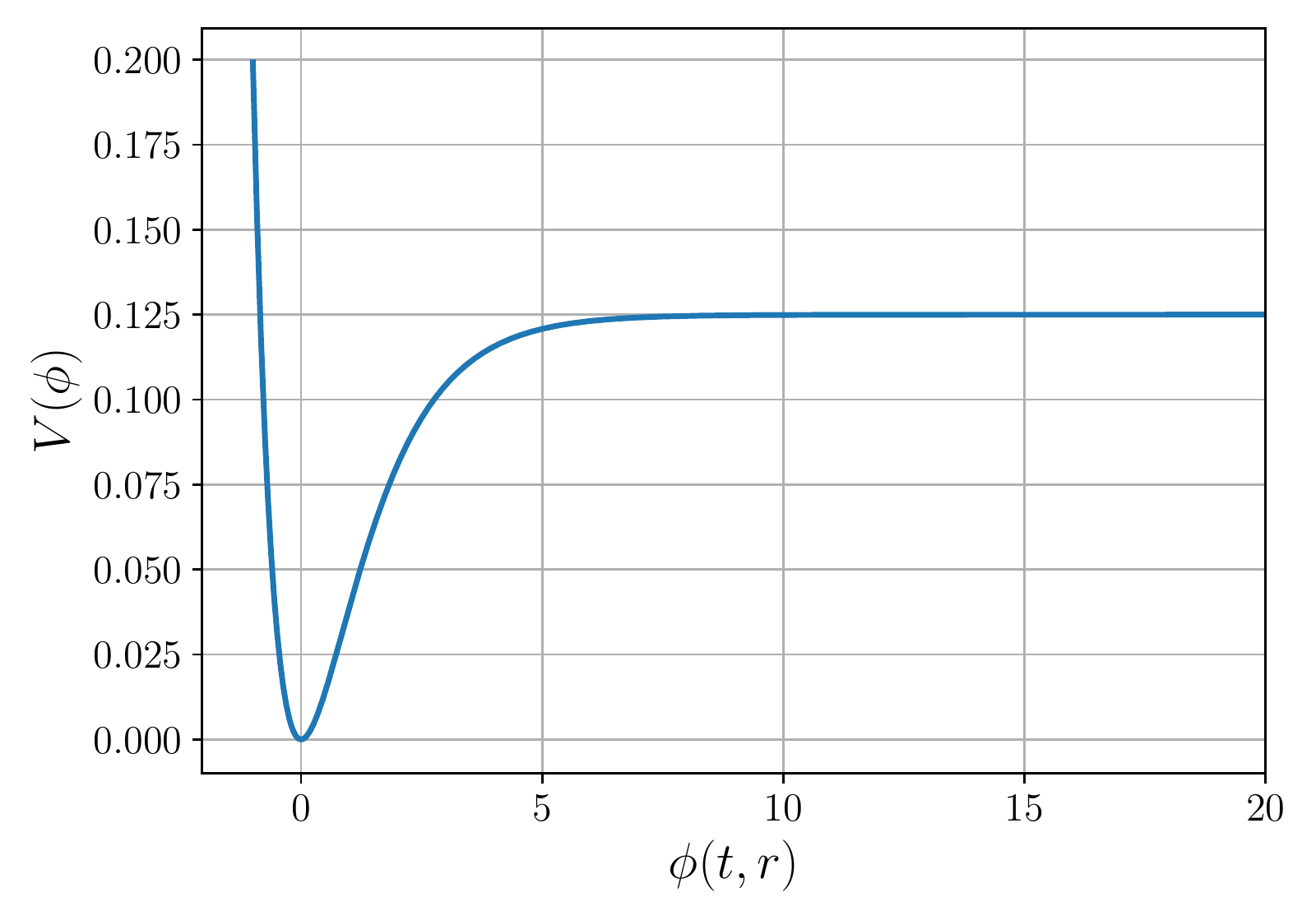}
    \caption{Potential for the Starobinsky $R^2$ model, $f(R) = R + R^2$, in the Einstein frame.}
    \label{fig:P_Starobinsky_model}
\end{figure}

In this model, the Ricci scalar in terms of the gravity scalar field is given by $R = ({\chi-1})/{2}$, and the potential (\ref{eq:fR_potential}) can be written as
\begin{equation}
    V(\phi) = \frac{1}{8} \left(1 - e^{-\sqrt{\frac{2}{3}} \phi}\right)^2
\end{equation}
and is plotted in Fig. \ref{fig:P_Starobinsky_model}. As we can see, $V(\phi)$ is constant at large $\phi$ and has a minimum at $\phi=0$; based on this behavior, inflation is a natural phenomenon in the $R^2$ model \cite{Maeda:1987xf}. \\

In this paper, we work in units in which $\kappa = 1$.
The system is evolved between $r=10^{-50}$ and $r = r_{max} = 50$ from $t=0$ until it forms an apparent horizon ($r_H$). 
We solve numerically the system of PDEs using the fourth-order Runge-Kutta method in both time and spatial directions. We work with a fixed grid in which the $r$-spacing is given by 
$\Delta r=10^{-4}$. 
The time resolution is also fixed but satisfies the Courant-Friedrichs-Lewy condition $\Delta t=\Delta r/5$.

\subsection{Weak field regime}
\begin{figure}[t!]
    \centering
    \begin{tabular}{ccc}
        \includegraphics[width=0.3\textwidth]{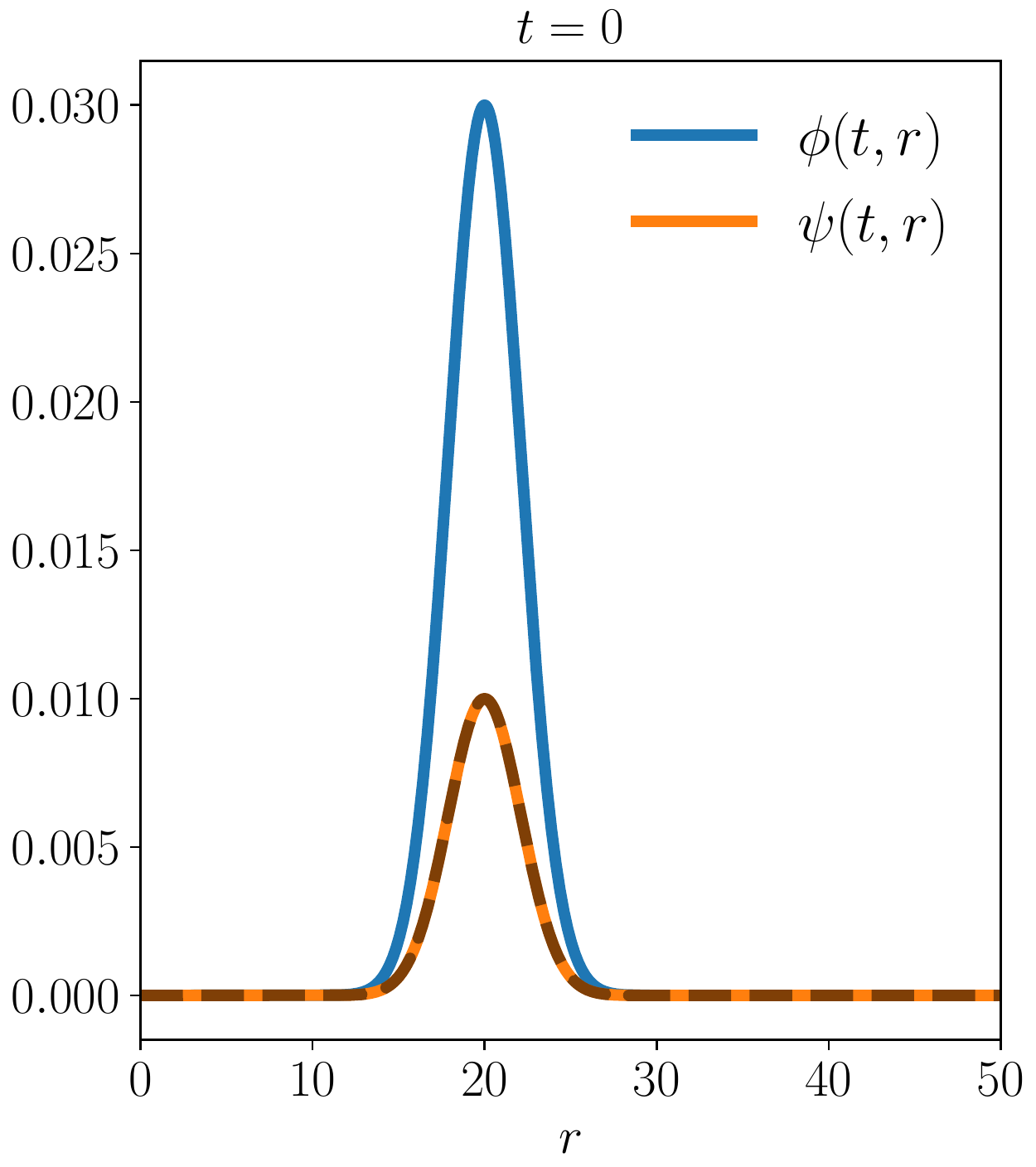} & 
        \includegraphics[width=0.3\textwidth]{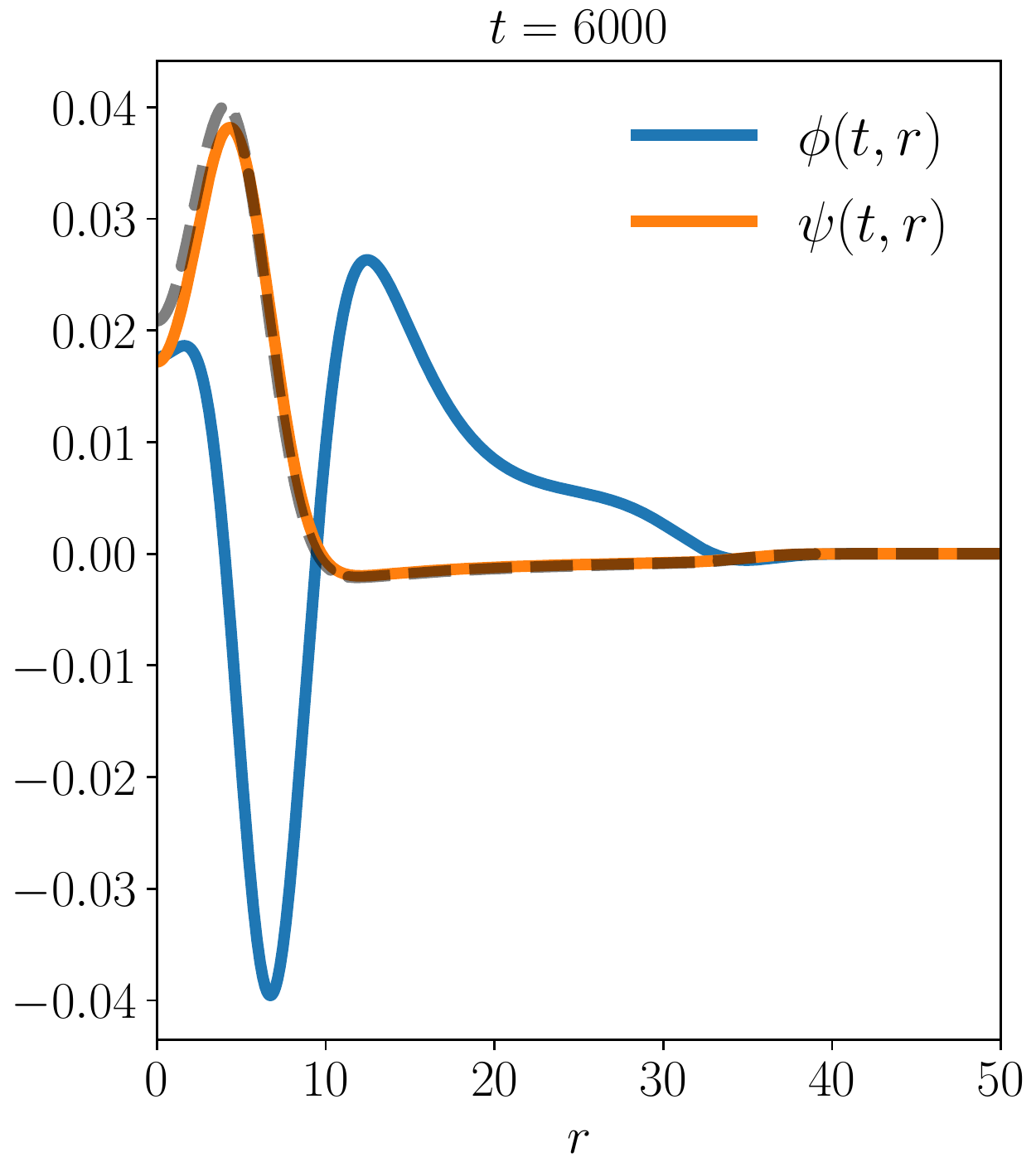} & 
        \includegraphics[width=0.3\textwidth]{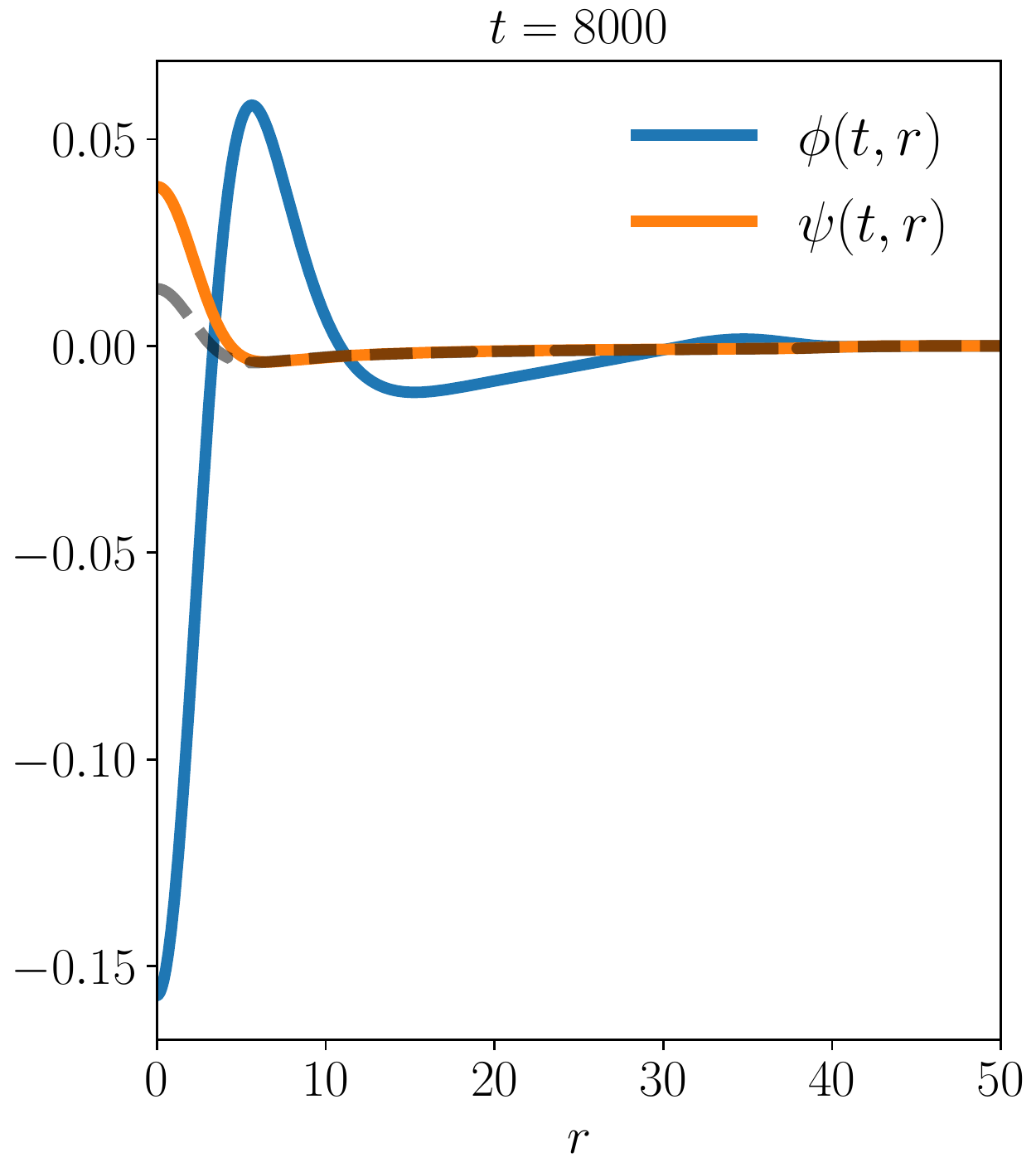} \\
        \includegraphics[width=0.3\textwidth]{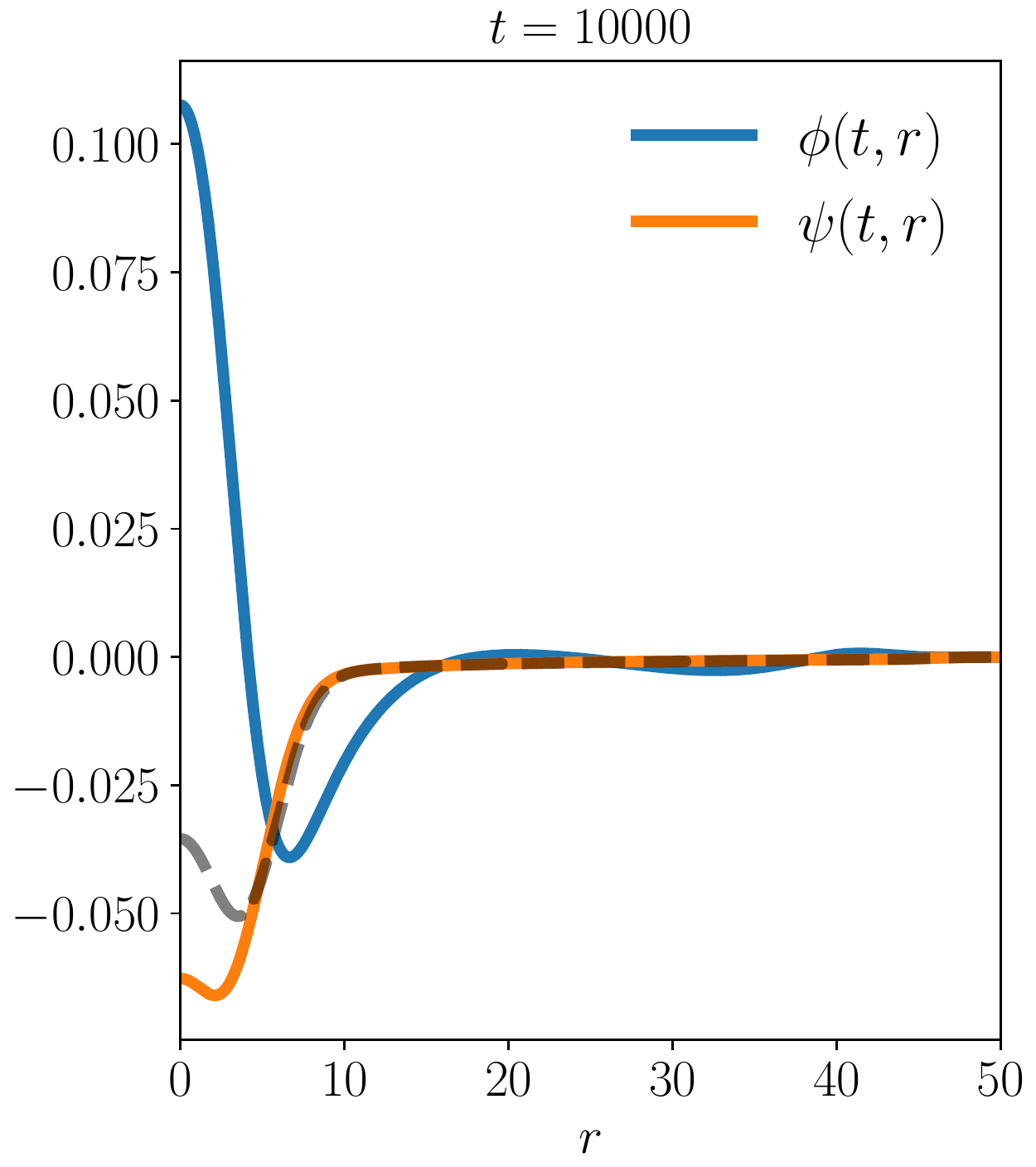} & 
        \includegraphics[width=0.3\textwidth]{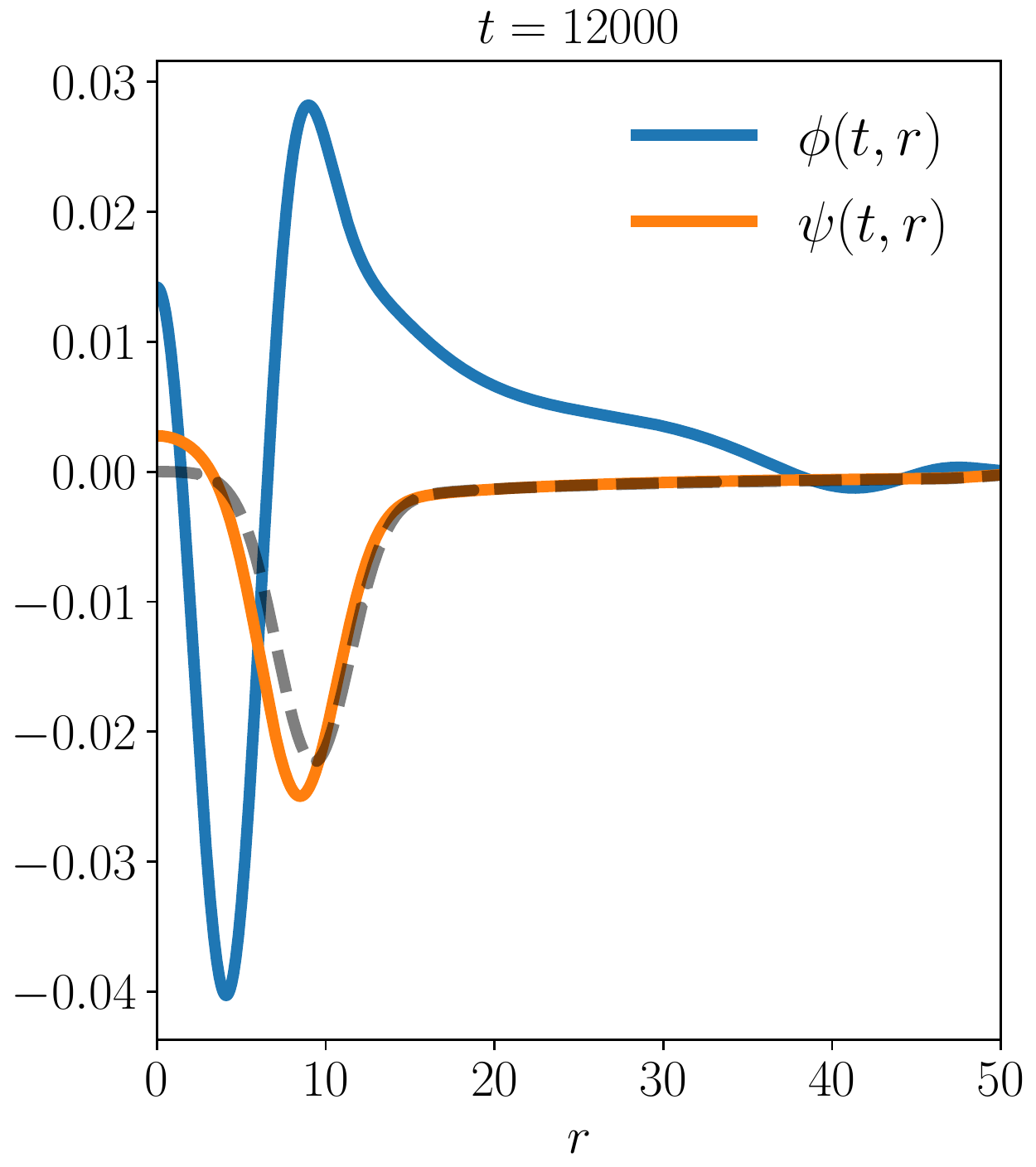} & 
        \includegraphics[width=0.3\textwidth]{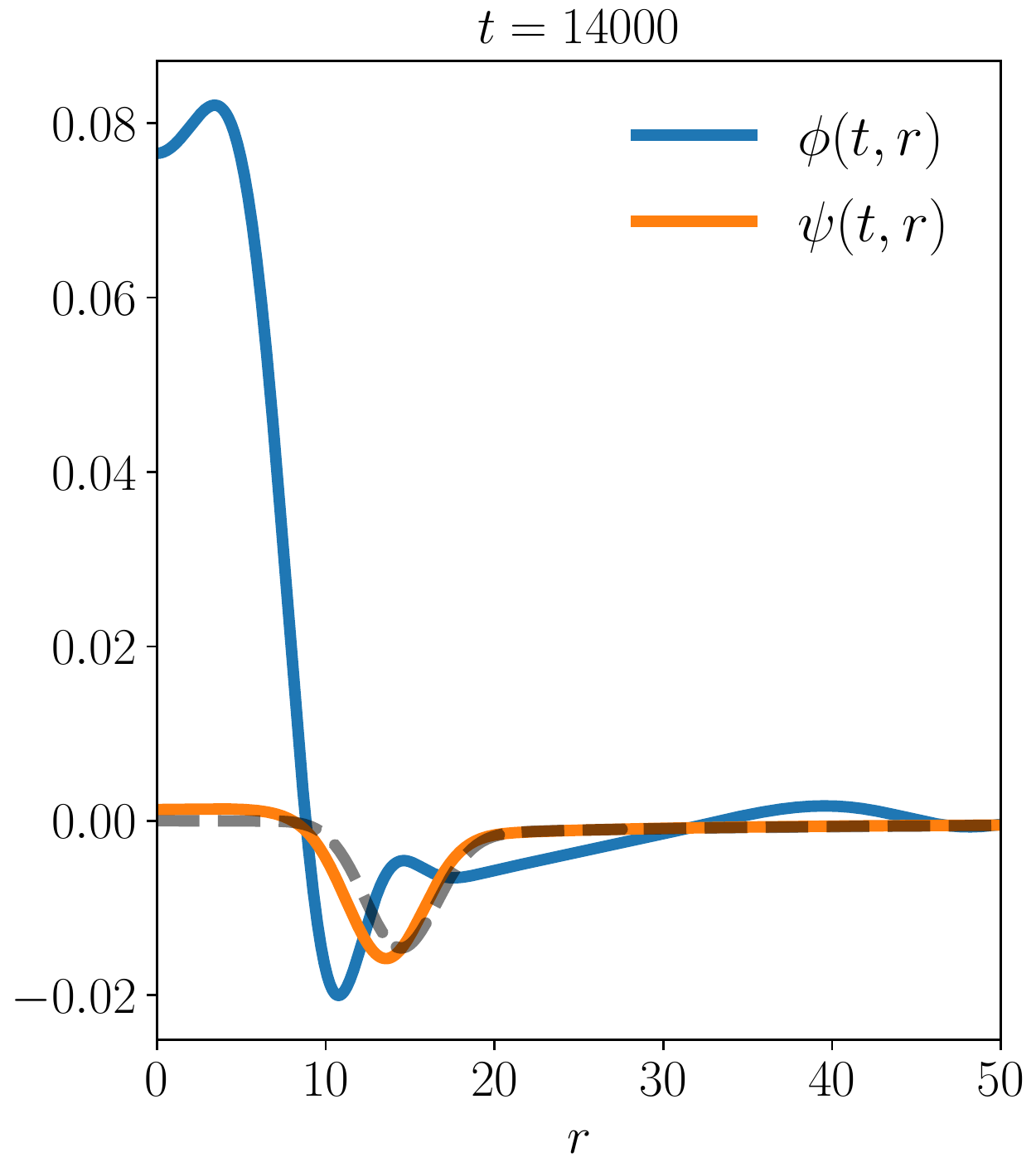} 
    \end{tabular}
    \caption{Plots of the scalar field profiles (in the weak field regime) at different times of the evolution. $\phi(t,r)$ (blue line) is the gravity scalar field. $\psi(t,r)$ (orange line) is the physical scalar field. The dashed line corresponds to the evolution of GR, i.e., the Choptuik gravitational collapse.}
    \label{fig:plot_fields_weak_limit}
\end{figure}

When the amplitude of the matter field and the amplitude of the gravity scalar field are small, $p_o^{(\phi)} \ll 1$ and $p_o^{(\psi)} \ll 1$, the gravity interaction is weak, and the matter field bounces at the origin $(r=0)$ and then disperses to infinity. 
In Fig. \ref{fig:plot_fields_weak_limit}, we plot both scalar fields at different time\footnote{Time is the iteration step and not proper time at $r=0$} for an evolution in which Eq. gives the initial data (\ref{eq:init_data}), and we fix $p_o^{(\psi)} = 10^{-2}$ and $p_o^{(\phi)} = 3\times 10^{-2}$. In this figure, we also plot the evolution for the Choptuik case [$f(R)$ model (\ref{eq::fR_model}) with $a = 0$].

We see that potential gravity does not capture enough matter to form a black hole, and after the interaction near the origin, the matter field is dispersed to infinity. 
In the same figure, we can notice that the inclusion of the term $R^2$ does not significantly modify the gravitational collapse compared to GR, the Choptuik gravitational collapse.
When $a \neq 0$, the matter field takes more time to bounce and therefore reaches infinity a bit later compared to the Choptuik case. This effect is no longer appreciated for smaller values of $p_o^{(\phi)}$ and $p_o^{(\psi)}$, in which we can conclude that the collapse is given by GR.

\subsection{Strong field regime}

It is expected that when we increase the initial mass of the matter field ($\psi$) added in the spacetime, the collapse of that field produces a black hole.
Also, if we fix the matter field amplitude $\qty(p_o^{(\psi)})$ but increase the amplitude of the gravity scalar field $\qty(p_o^{(\phi)})$, it is expected that potential gravity becomes stronger, also forming a BH. 

\begin{figure}
    \centering
    \includegraphics[scale=0.7]{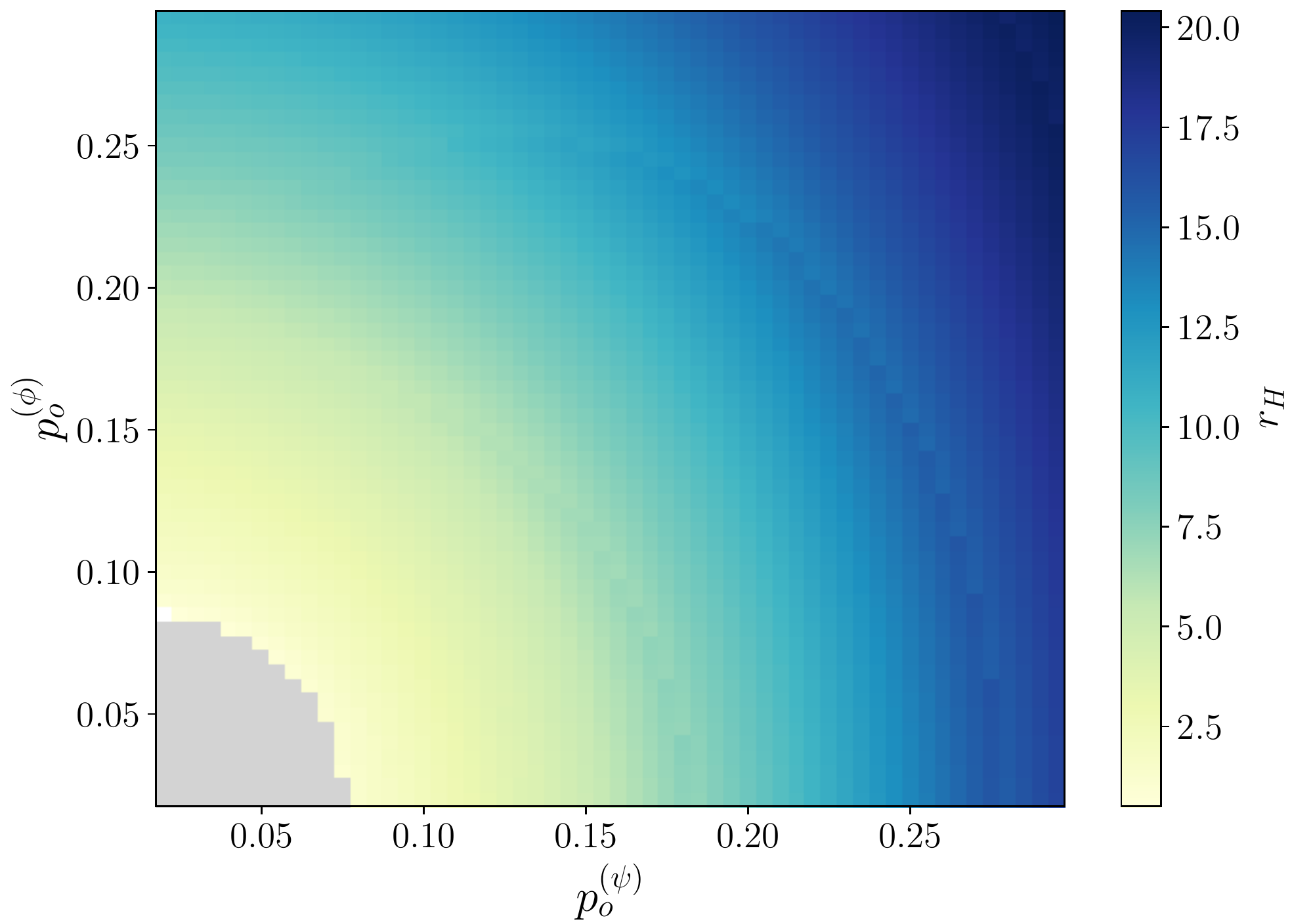}
    \caption{Radius of the formed BH ($r_h$) depending on the value for the initial data $\qty(p_o^{(\psi)}, \; p_o^{(\phi)})$.}
    \label{fig:radius_formation}
\end{figure}

In this paper, we run simulations in an extended parameter space $\qty(p_o^{(\psi)}, \; p_o^{(\phi)})$ from values in which all fields are dispersed to infinity, to values where the final result of the evolution is the creation of an event horizon.
In Fig. \ref{fig:radius_formation}, we plot the radius of the formed BH at the end of each evolution for the parameter space explored in this paper. 
It can be noticed that for small values of both scalar fields, there is no BH formation.
When we increase one of the two amplitude parameters, inevitably, the evolution stops, signaling a formation of a BH. This is what we anticipated since, for large values of $p_o^{(\psi)}$, we add more matter at the initial time, while for large values of $p_o^{(\phi)}$, the gravitational potential captures more matter and forms the BH. 
As expected, the radius of the BH starts from small values and increases while the initial mass of the fields added to the spacetime becomes stronger. 

As we saw in Fig. \ref{fig:radius_formation}, a BH is always formed once we cross the critical surface. 
However, Ref. \cite{Zhang:2016kzg} found that there are initial data that rely on the BH formation side that does not produce a BH. 
As shown in Fig. 3 of that paper, these patches in the parameter space are small regions that probably are not captured by our discretization of the parameter space.\footnote{It can also be mentioned that the last term in Eq. (13) of Ref. \cite{Zhang:2016kzg} seems to have the wrong sign. It remains to be checked if it is a typo.}

Although we do not show it here, every time a black hole forms, the curvature scalar $R$ and the Ricci tensor squared $R_{\mu\nu} R^{\mu\nu}$ are zero at the last time of the evolution and outside the event horizon, $r > r_H$; thus the exterior solution is Schwarzschild, which is consistent with the no-hair theorem and previous results showed in \cite{WHITT1984176}.

It can be mentioned that the critical collapse of the Choptuik model is of type-II, which means that the radius of the BH near the critical limit ($p^\star$) approaches zero; thus, through a fine-tuning mechanism, we can produce infinitesimal BHs.
For the Starobinsky model, the radius of the formed BH near the critical limit seems to start at a finite value but close to zero (near the critical limit $r_H \approx 10^{-2}$).

Performing a fine-tuning method between the weak-field regime and the strong-field regime, we found the critical surface. Near this limit, the radius of the BH satisfies a scaling behavior
\begin{align}
r_H\propto (p-p^\star)^{\gamma} \, ,
\end{align}
similar tho the Choptuik case \cite{Choptuik:1992jv}.
Our results show that by fixing $p_o^{(\phi)}$ and varying the amplitude of the matter scalar field $p_o^{(\psi)}$, we found that $\gamma \approx 0.38$. However, by setting the amplitude of the initial matter scalar field to zero to study vacuum collapse in the Starobinsky model, we obtain $\gamma \approx 0.50$.
We are performing more detailed calculations to get closer to the critical solution in this example and to get more details on the power-law scaling.

\section{Conclusions and discussions} 

In this paper, we have studied gravitational collapse for the Starobinsky $R^2$ model (a particular example of $f(R)$ theory) coupled to a massless scalar field in a spherically symmetric spacetime. 
The Starobinsky model has been used as a possible explanation for inflation in the early universe. Also, it is considered the first quantum correction to the vacuum Einstein equations and was expected to avoid the singularity problem at the center of black holes. 

We have presented the framework of $f(R)$ theory in the Jordan frame and performed a conformal transformation to write this theory as a Scalar-Tensor theory. In the new frame, called the Einstein frame, besides the standard matter field added to the spacetime, gravity is coupled to a new scalar field which drives the modifications from the Einstein gravity.

The gravitational collapse for the Starobinsky $R^2$ model is very similar to the Choptuik cases [$f(R) = R$], in the sense that depending on the initial conditions, the final state of the system is the formation of a BH or dispersion of the matter field to infinity. 
Our results show that in a spherically symmetric spacetime, when the amplitude of the initial matter scalar field and the gravity scalar field are weak, the final result of the evolution is a Minkowski spacetime; the matter field is dispersed to infinity. A BH is formed when we increase the amplitude of one of the scalar fields. 
Increasing the matter field amplitude adds more mass to the spacetime at the initial time; while increasing the amplitude of the gravity scalar field, the gravitational potential capture more matter, forming a BH in both cases. 
When a BH is formed, the final solution is the Schwarzschild metric. These results are consistent with those obtained in Ref. \cite{WHITT1984176}.

Finally, analyzing the critical phenomena near the formation of the event horizon, we found that in the critical limit, a power-law scaling for $r_H$ appears with a critical exponent of $\gamma \approx 0.38$. 
We are analyzing the case of spherical vacuum collapse in the Starobinsky model with more detailed evolution since the critical exponent is different. 

\bigskip

{\bf Acknowledgment:} The author would like to thank VRIEA-PUCV. All simulations were executed on the {\it Chaska Cluster} at Instituto de Física, PUCV.

\spacing{1}
\providecommand{\href}[2]{#2}\begingroup\raggedright\endgroup

\end{document}